\documentclass[preprint,12pt]{elsarticle}

\usepackage{amssymb}
\usepackage{amsmath}
\usepackage{lineno}
\usepackage{amsmath}
\usepackage{lineno} 
\usepackage{bm}
\usepackage{multirow}
\usepackage{algorithm}
\usepackage{algpseudocode}
\usepackage{xcolor}

\begin{document}

\begin{frontmatter}

%% Title, authors and addresses

%% use the tnoteref command within \title for footnotes;
%% use the tnotetext command for theassociated footnote;
%% use the fnref command within \author or \affiliation for footnotes;
%% use the fntext command for theassociated footnote;
%% use the corref command within \author for corresponding author footnotes;
%% use the cortext command for theassociated footnote;
%% use the ead command for the email address,
%% and the form \ead[url] for the home page:
%% \title{Title\tnoteref{label1}}
%% \tnotetext[label1]{}
%% \author{Name\corref{cor1}\fnref{label2}}
%% \ead{email address}
%% \ead[url]{home page}
%% \fntext[label2]{}
%% \cortext[cor1]{}
%% \affiliation{organization={},
%%             addressline={},
%%             city={},
%%             postcode={},
%%             state={},
%%             country={}}
%% \fntext[label3]{}

\title{Human Gaze-based Dual Teacher Guidance Learning for Semi-Supervised Medical Image Segmentation}

%% use optional labels to link authors explicitly to addresses:
\author[1]{Rongjun Ge}
\author[2]{Chong Wang}
\author[4]{Yuxin Liu}
\author[5]{Chunqiang Lu}
\author[6]{Cong Xia}
\author[7]{Yehui Jiang}
\author[8]{Fangyi Xu}
\author[6]{Yinsu Zhu}
\author[2]{Daoqiang Zhang}
\author[1]{Chengyu Liu\corref{cor}} 
\ead{chengyu@seu.edu.cn}
\author[4]{Yang Chen\corref{cor}} 
\ead{chenyang.list@seu.edu.cn}
\author[3]{Shuo Li} 
\author[3]{Yuting He\corref{cor}}
\ead{yuting.he4@case.edu}
\cortext[cor]{Corresponding author} 
 \affiliation[1]{organization={School of Instrument Science and Engineering, Southeast University},
             city={Nanjing},
             postcode={210096},
             country={China}}
 \affiliation[2]{organization={College of Artificial Intelligence, Nanjing University of Aeronautics and Astronautics},
             city={Nanjing},
             postcode={211106},
             country={China}}

  \affiliation[4]{organization={School of Computer Science and Engineering, Southeast University},
             city={Nanjing},
             postcode={210096},
             country={China}}
 \affiliation[5]{organization={Nurturing Center of Jiangsu Province for State Laboratory of AI Imaging and Interventional Radiology, Department of Radiology, Zhongda Hospital, Medical School of Southeast University},
             city={Nanjing},
             postcode={210009},
             country={China}}
             
 \affiliation[6]{organization={Department of Radiology, The Affiliated Cancer Hospital of Nanjing Medical University, Jiangsu Cancer Hospital, Jiangsu Institute of Cancer Research},
             city={Nanjing},
             postcode={210000},
             country={China}}

 \affiliation[7]{organization={Department of Ultrasound Medicine, the Second Affiliated Hospital of Anhui Medical University},
             city={Hefei},
             postcode={230601},
             country={China}}

 \affiliation[8]{organization={Department of Radiology, Sir Run Run Shaw Hospital, School of Medicine, Zhejiang University},
             city={Hangzhou},
             postcode={310016},
             country={China}}

\affiliation[3]{organization={School of Medicine, Case Western Reserve University},
             city={Cleveland},
             country={USA}}

%% Abstract
\begin{abstract}
%% Text of abstract
{In the field of medical image segmentation, the scarcity of labeled data poses a major challenge for existing models to accurately perceive target regions. Compared with manual annotation, gaze data is easier and cheaper to obtain. {As a classical semi-supervised learning framework, mean-teacher can effectively use a large number of unlabeled medical images for stable training through self-teaching and collaborative optimization.} 
Our study is based on the mean-teacher framework. 
{By combining gaze data, it aims to address two crucial issues in semi-supervised medical image segmentation: 1) expand the scale and diversity of the dataset with limited labeled data; 2) enhance the network's perception ability.} 
We propose the Human Gaze-based Dual Teacher Guidance Learning model (HG-DTGL). In this model, human gaze serves as an additional hidden ‘teacher’ in the mean-teacher architecture. We introduce the GazeMix to generate reliable mixed data to expand the diversity and scale of the dataset, and the Multi-scale Gaze Perception (MGP) module is used to extract the multi-scale perception of the network. A Gaze Loss is designed to align the model's perception with human gaze. We have verified HG-DTGL on multiple datasets of different modalities and achieved superior performance on a total of ten different organs/tissues, with extensive experiments. This demonstrates that our method has strong generalization ability for medical images of different modalities, and shows the great application potential of gaze data in semi-supervised medical image segmentation.
}
\end{abstract}

%% Keywords
\begin{keyword}
%% keywords here, in the form: keyword \sep keyword

%% PACS codes here, in the form: \PACS code \sep code

%% MSC codes here, in the form: \MSC code \sep code
%% or \MSC[2008] code \sep code (2000 is the default)
Human Gaze \sep Semi-Supervised \sep Medical Image Segmentation
\end{keyword}

\end{frontmatter}

%% Add \usepackage{lineno} before \begin{document} and uncomment 
%% following line to enable line numbers
%% \linenumbers

%% main text
%%

%\linenumbers
\section{Introduction}
\label{sec:intro}
Semi-supervised segmentation \cite{sohn2020fixmatch, miyato2018virtual, li2020transformation, lai2021semi} has garnered heightened attention in recent years and has become a pervasive trend in the realm of medical image analysis. In the realm of semi-supervised medical image segmentation, existing methods usually improve the segmentation performance by changing the model structure\cite{li2020transformation} and consistency regularization \cite{sohn2020fixmatch, xie2020unsupervised}. However, {the challenge primarily comes from} the limited quantity of labeled data, making it difficult to accurately estimate the distribution from this dataset. To expand the dataset's size and diversity, weak data augmentation techniques like rotation and flipping are frequently applied. Additionally, MixUp \cite{zhang2017mixup} and CutMix \cite{yun2019cutmix} stand as robust data processing methods that have the potential to encourage unlabeled data to assimilate common semantics from the labeled data. This is because pixels within the same map share semantics and tend to be closely related \cite{wang2022separated}. However, most existing CutMix-based methods focus solely on applying CutMix across unlabeled data or merely copying segments from labeled data to serve as foreground and pasting them onto other data. CutMixed images are cropped randomly or centered on the image, which can lead to the loss of segmented targets. Therefore, considering the use of guidance information to assist generation, the usability of the mixed images can be ensured. 

\begin{figure}[!htb]
    \centering
    \includegraphics[width=0.7\linewidth]{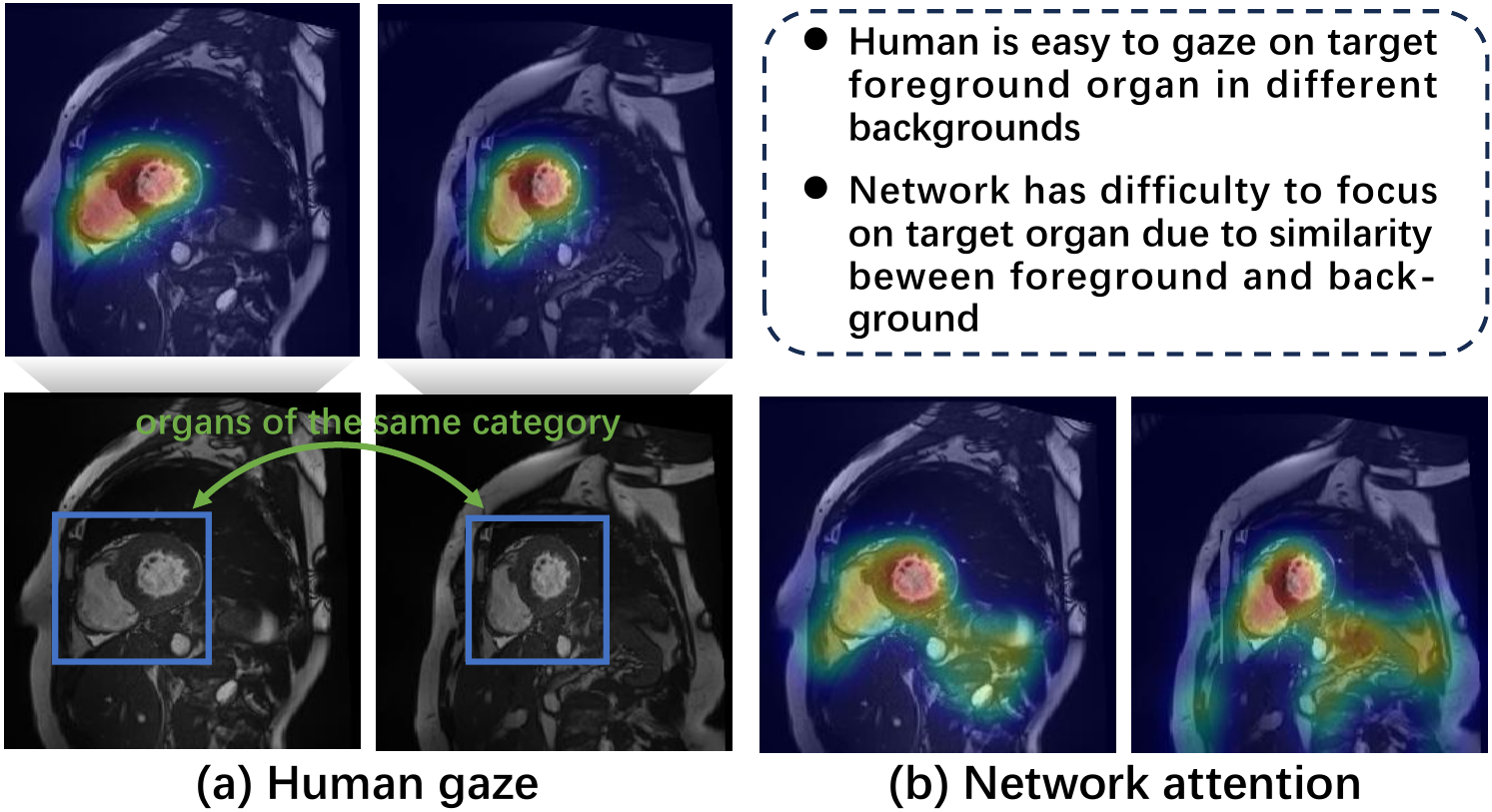}
    \caption{ {\textbf{(a)
    Human gaze heatmap:} Humans are} able to accurately gaze and identify the correct foreground organ region even with various backgrounds. 
    {\textbf{(b)  
    Network attention heatmap:} Due to the similarity between the segmentation foreground and background in medical images, the network faces challenges in focusing on the target.}  }
    \label{mix}
\end{figure}

Human beings possess the ability to accurately fixate on the same target even when the surrounding background undergoes changes. As shown in Figure \ref{mix}, humans are able to accurately gaze and identify the correct foreground organ region even with various backgrounds, which is difficult for the network to {pay attention to the same target} due to interference from background and irrelevant
structure. The gaze information obtained from radiologists provides valuable human perceptual insights \cite{cheng2022eye, narganes2023explicit}, which can be leveraged to guide the generation of effective mixed images. Eye gaze data is {insensibly gathered during the radiologist's interpretation}, without any disruption to their diagnostic workflow and additional time cost. 
Typically, eye gaze data aptly reflects the spatial orientation of the target to be segmented, thereby facilitating precise target localization and enabling the implementation of effective mix strategies. 
Furthermore, this gaze information can be leveraged to generate heatmaps, which provides insights into the areas of human attention directed toward the target. Utilizing heatmaps to constrain network spatial perception areas effectively adjusts the spatial feature of network output, and enhances network segmentation performance.

{By integrating the gaze data of radiologists, the model is able to further adjust the segmented target area to strengthen the model's sensitivity to complex lesions and improve the diagnostic accuracy of difficult and complicated diseases in clinical practice. For example, in the early clinical diagnosis of lung cancer, tiny or early-stage lesions are likely to be overlooked by traditional segmentation models. Khosravan et al.\cite{khosravan2017gaze2segment} proposed a multimodal attention fusion mechanism named Gaze2Segment. It employs an eye-tracker to capture radiologists' diagnostic gaze data during chest CT reading, converts gaze data into heatmaps that are computationally fused with grayscale saliency features. Such synergistic integration enables the model to achieve sub-millimeter lesion detection accuracy, effectively bridging the gap between algorithmic segmentation and clinical diagnostic workflows.}

\begin{figure*}[htb]
    \centering
    \includegraphics[width=1\linewidth]{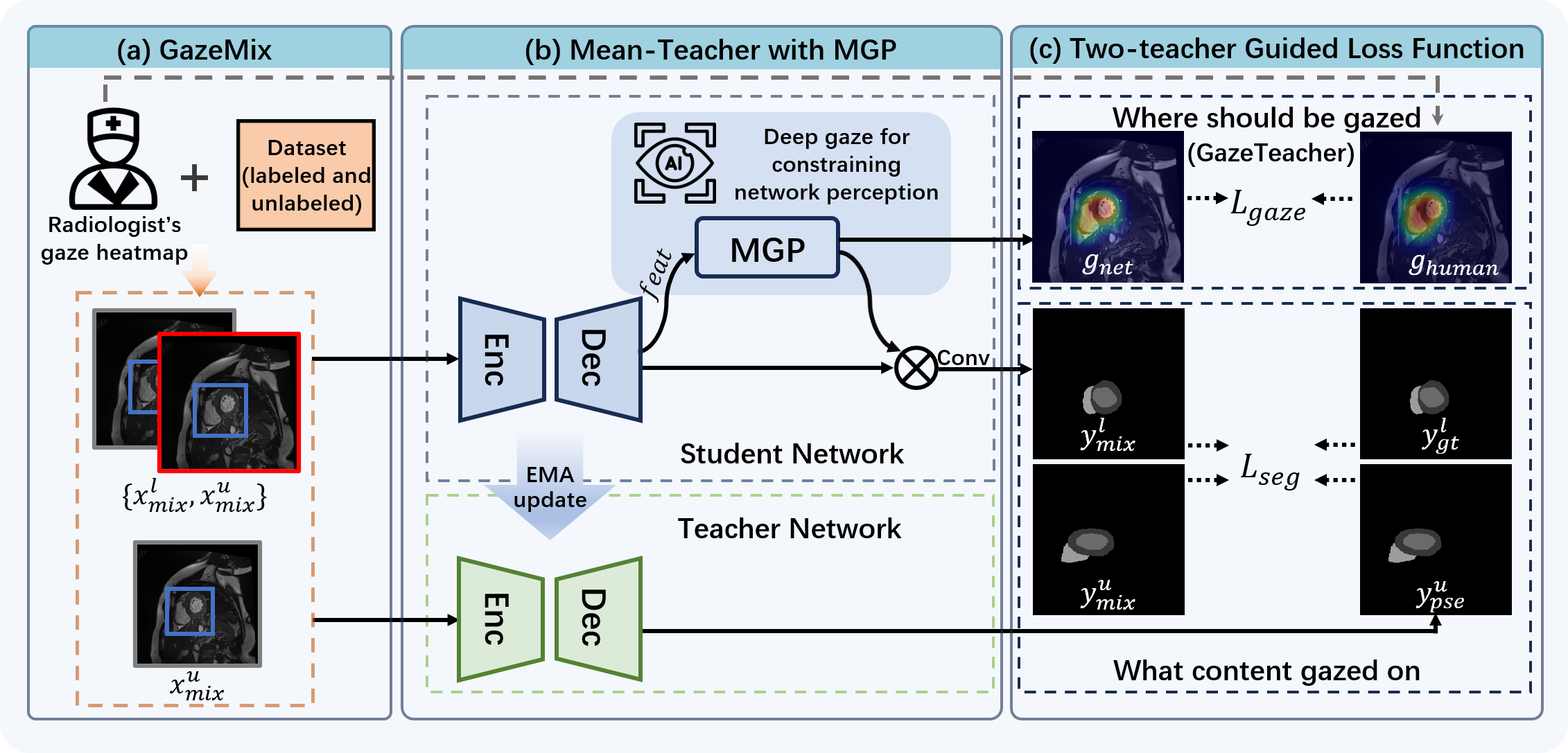}
\vspace{-0.2in}
    \caption{{Overall architecture of our proposed HG-DTGL model for semi-supervised medical image segmentation. (a) GazeMix combines the radiologist's gaze heatmap with a dataset containing both labeled and unlabeled data to generate mixed image pairs. (b) The Mean-Teacher network architecture with MGP includes encoder-decoder structures in both the student and teacher networks. MGP enhances feature capture ability, and the teacher network is updated via EMA to provide pseudo labels for the student network. (c) Two-teacher guided loss function constrains the model from two aspects: "where should be gazed" and "what content gaze on". } }
    \label{network}
\vspace{-0.1in}
\end{figure*}

\subsection{Related Work}
Traditional medical image segmentation networks \cite{ronneberger2015u, milletari2016v} are susceptible to interference from the background and irrelevant structures, which affects the network perception of target foreground.
Although self-attention based methods \cite{chen2021transunet,cao2022swin} are widely applied in the field of medical image segmentation today, these methods analyze the attention of the network based on global features, {and are unable to reflect the network's spatial attention towards the image}. CAM \cite{zhou2016learning} is typically used in classification tasks. It reflects the final result determined by the network. The result is presented in the form of a heatmap based on the regions and their importance levels. Besides, thanks to  SE-Net \cite{hu2018squeeze} and CBAM \cite{woo2018cbam}, networks can adaptively adjust their importance judgments for different regions and dimensions of input image features based on their perceptual capabilities.
However, in some situations where the foreground and background are blurred or similar, it is difficult for the network to distinguish between them. Therefore, human perceptual information can be used to constrain the network, thereby enhancing its perception ability for complex situations.

Semi-supervised medical image segmentation methods
can be roughly grouped into three categories. (1) Pseudo-labeling methods \cite{pseudo2013simple,dong2018tri,zhou2005tri,rizve2021defense}, which generate pseudo-labels for
unlabeled data as classic practice in semi-supervised learning. The key point of the pseudo-labeling method is how to
generate reliable pseudo-labels. 
(2) Consistency regularization-based methods \cite{bortsova2019semi,fang2020dmnet,li2020shape,luo2021semi,wang2022semi,xia2020uncertainty,yu2019uncertainty}, which attend to various levels of information for a single target via multi/dual-task learning or transformation consistent learning. 
(3) Contrastive learning-based methods \cite{wu2022exploring, you2022simcvd}, which learn representations that maximize the similarity among positive
pairs, and minimize the similarity among negative pairs.

Eye tracking has been utilized in medical image analysis. {Stember et al.}\cite{stember2020integrating} integrated eye tracking and speech recognition for accurate annotation of MR brain images. The generated eye-tracking masks are then used as input to convolutional  neural networks for segmentation, and the results are compared with those obtained by hand annotation. {Wang et al.}\cite{wang2022follow} considered gaze data as human visual attention and utilizes it to guide the network attention from class activation map (CAM) \cite{zhou2016learning} with an attention consistency module in osteoarthritis assessment of knees. 
And {Wang et al.}\cite{wang2023eye} constructed a dual-path parallel network to model the relationship between eye gaze data and medical images, with interaction between paths to improve performance.
In another work \cite{ma2023eye}, {Ma et al.} intended to incorporate experts' intelligence that gaze data can reveal, into shortcut learning of vision transformers. It demonstrates that the shortcuts can be effectively rectified by learning from the experts' domain knowledge. 
{Besides, recent advances in gaze-based augmentation methods have further demonstrated the value of visual attention patterns for enhancing model training. Wang et al.\cite{wang2025learning} proposed gaze-conditioned augmentation by modulating image transformations according to fixation density maps, so that benefited the medical image classification. Zhao et al.\cite{zhao2024mining} embedded the gaze data into the contrastive learning framework to generate positive sample pairs through gaze similarity, so that enabled the model to learn semantic representations that are in line with clinical decision-making.}
As can be seen, the use of eye gaze information has great value and potential in medical image analysis.

\subsection{Overview of the Proposed Model}
To this end, we propose a novel Human Gaze-based Dual Teacher Guidance Learning model (HG-DTGL), which is built in mean-teacher network architecture and incorporates human gaze as an additional hidden teacher for semi-supervised medical image segmentation. As shown in Figure \ref{network}, gaze ‘teacher’ guides HG-DTGL model to learn ``where should be gazed'' with human gaze perception,  
and network ‘teacher’ drives learning on ``what content gazed on'' with its pseudo-label.
Specifically, based on human gaze, we design a new data augmentation strategy GazeMix to greatly reduce the reliance on large amounts of annotated data and encourage unlabeled data to assimilate common semantics. To encourage the model to learn rich semantic information, the crop patch of human gaze in the labeled and unlabeled images is cropped. Then, it is pasted on another unlabeled image, generating new mixed image pairs for student network. The GazeMix {is consistently applied to the images}, labels, and human gaze heatmaps. It incorporates complete target information and boundaries which mitigate the edge crop of traditional copy-paste strategy. Furthermore, to alleviate the empirical difference between network perception and human perception, we have designed a Multi-scale Gaze Perception (MGP) module. It simultaneously considers both fine-grained details and holistic features, enabling complementary information from different scales, thereby reducing errors introduced by a single scale. Human gaze perception, as a hidden teacher, constitutes the Gaze Loss with network perception and guides the correct region of network gaze to align with human gaze. Furthermore, the pseudo labels predicted by the teacher network are used to supervise unlabeled data. 

The main contributions are summarized as: 
\textbf{1)}
For the first time, a dual teacher guidance learning architecture stimulated by human gaze is proposed for semi-supervised medical image segmentation, 
in which gaze ‘teacher’ enables student network understanding “where should be gazed” with human gaze perception, {and the network ‘teacher’ guides the student network to learn} “what content gazed on” through massive pseudo labels. 
Thus, the model is able to focus and deeply absorb foreground knowledge from rich unlabeled data, breaking the shackles of limited labels.
\textbf{2)} 
Based on human gaze, we design a novel data augmentation strategy GazeMix with labeled and unlabeled images. It greatly expands semantic feature space of dataset, reduces reliance on large amounts of annotated data, and encourages unlabeled data to assimilate common semantics. 
\textbf{3)}
We newly propose a Multi-scale Gaze Perception module (MGP) and Gaze Loss to {teach the model to align with humans and learn by focusing on critical regions like radiologist}. MGP comprehensively extracts multi-scale network perception on medical image covering fine-grained and holistic structures, 
whereas Gaze Loss is employed to narrow the gap between human gaze and network perception. This strategic integration bolsters the network's segmentation performance through cognition alignment between human and network. 
\textbf{4)} Extensive experiments are implemented to systematically evaluate model performance and deeply analyze effects of components by comparing with SoTA methods, on public datasets ACDC \cite{bernard2018deep}, CAMUS \cite{leclerc2019deep}, Synapse \cite{landman2015miccai} and SCR \cite{shiraishi2000development}.

\section{Methodology}
\label{sec:method}
As shown in Figure \ref{network}, HG-DTGL is built in mean-teacher network architecture and incorporates human gaze as an additional hidden ‘teacher’. 
1) The gaze ‘teacher’ works through the collaboration of MGP and Gaze Loss $\mathcal{L}_{gaze}$ to teach student network “where should be gazed”, driving alignment between network perception and human gaze. 
2) The network ‘teacher’ works with the GazeMix to teach student network  “what content gazed on”, {through the supervision of pseudo labels for unlabeled data}, and the ground truth supervises labeled data. The teacher network is also pre-trained with GazeMix.

\begin{figure}[t]
    \centering
    \includegraphics[width=0.75\linewidth]{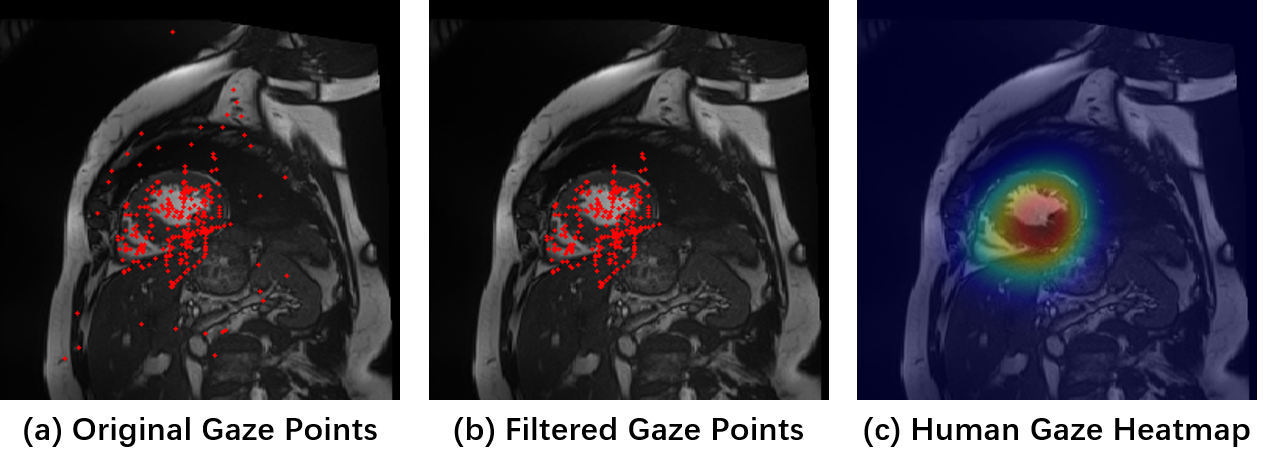}
        \caption{Gaze points before (a) and after (b) the saccades filtered out.
The generated Human Gaze Heatmap with filtered gaze points clearly reflects the location and distribution 
that human precisely perceives. The red
points in (a) and (b) are gaze points drawn on the corresponding image.}
\label{points}
\vspace{-0.05in}
\end{figure}

\subsection{Gaze Data and Gaze Movement Heatmap}
{We have designed a user-friendly system to collect eye-tracking data during radiologists reading medical images. In order to ensure data accuracy and scalability, we have set two constraints: 1) The collection frequency of the eye tracker is set to 60 gaze points per second; 2) The upper limit of the collection time for a single sample is 20 seconds. This design enables the gaze points to be controlled within the range of 800-1200 per sample. Additionally, we have developed a gaze data collection process guide for radiologists, and they collected the gaze data in accordance with this guide. The gaze data collection process is in the Supplementary Materials. 
 } 

In the subsequent processing of gaze data, we classify the gaze points into two categories: fixations and saccades. Fixation points correspond to the areas where the radiologist focuses attention during the screening process, while saccade points represent rapid eye movements that occur during the transitions between fixations. In order to precisely perceive radiologist’s visualized concern in imaging reading, the fixation points are further extracted from gaze points with saccades filtered out. To identify the saccade points, we use the timestamps to calculate the velocity of eye movements and set a threshold value for the maximum allowable velocity. Any gaze points exceeding this threshold are classified as saccades and excluded. By filtering out the saccade points, the range that radiologists actually pay attention to during the screening process is obtained, as:
 
\begin{equation}
\small
p_{i}\!=\!
\begin{cases}
P_{sac}\;\;\; \text{if}\; \frac{\left\|p_{i}-p_{i-1}\right\|_{2}}
{\Delta d} >v_{th}\;\&\;\frac{\left\|p_{i}-p_{i+1}\right\|_{2}}{\Delta d} >v_{t h}, \\
P_{fix}\;\;\; \text{if}\;\frac{\left\|p_{i}-p_{i-1}\right\|_{2}}
{\Delta d} <v_{th}\;\&\; \frac{\left\|p_{i}-p_{i+1}\right\|_{2}}{\Delta d} <v_{t h},
\end{cases}
\end{equation}  
where $p_{i}$ denotes gaze point, $P_{\text {sac}}$ and $P_{\text {fix}}$ are the sets of
saccade points and fixation points, $\Delta d$ is the duration between the gaze points,
and $v_{t h}$ means the velocity threshold.
Following the filtering process, human gaze heatmap is generated with the density of gaze points distribution on the medical image. {As shown in Figure \ref{points}}, the human gaze heatmap with saccade points filtered out reflects the location and distribution that human precisely perceives.

\begin{figure}[tp]
    \centering
    \includegraphics[width=0.55\linewidth]{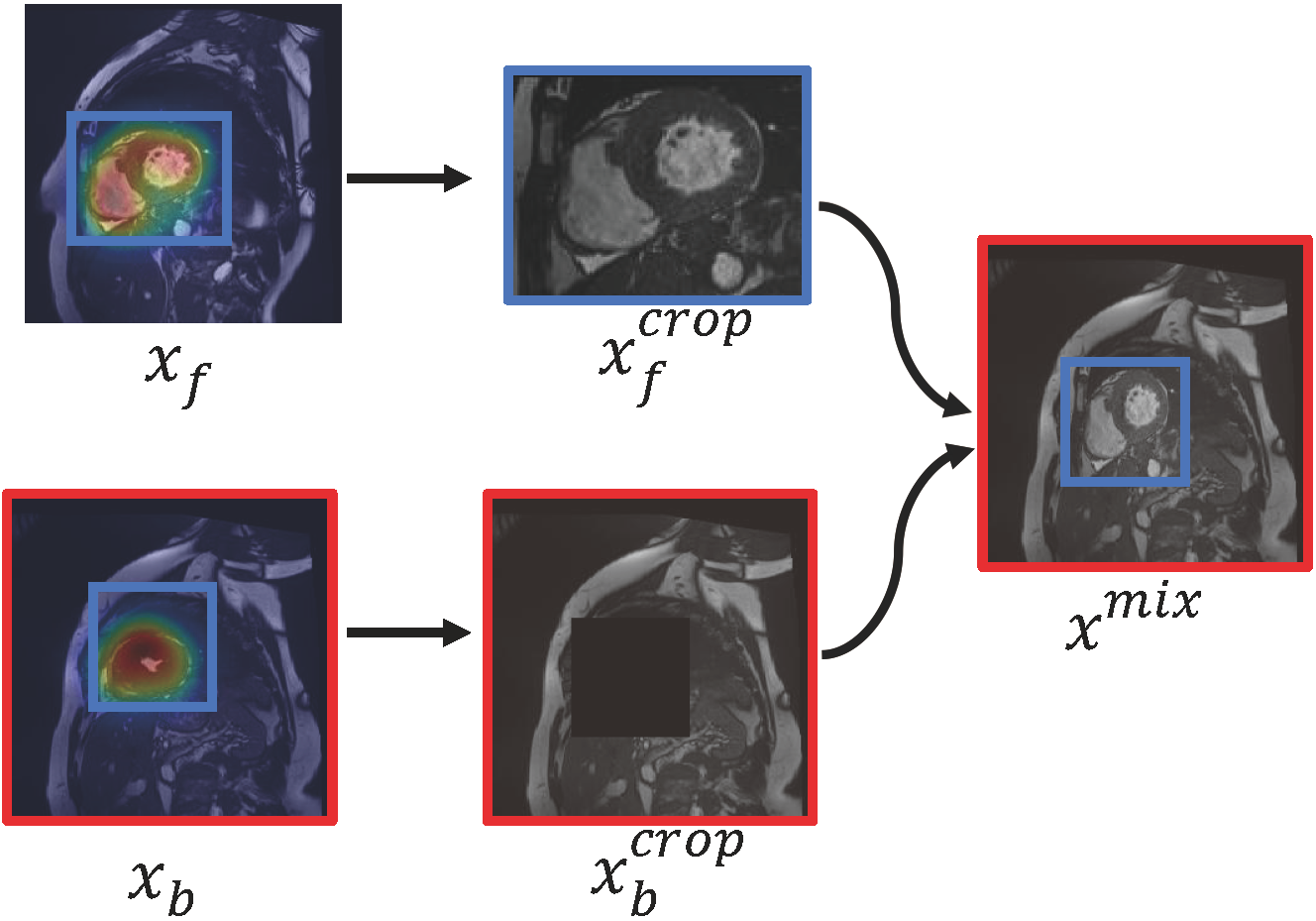}
\vspace{-0.15in}
    \caption{Illustration of the human visual perception-based GazeMix procedure.}
\label{gazemix}
\vspace{-0.2in}
\end{figure}

\subsection{GazeMix}
GazeMix is a novel data augmentation technique that improves model robustness with human visual perception, and slots in easily to existing training pipelines. The pipeline of the proposed GazeMix method is shown in Figure \ref{gazemix}.

We define medical image as $ x \in \mathbb{R}^{W \times H}$, and the per-pixel label $y \in\left \{ 0,1,...,K-1 \right \} ^{W \times H}$, where $K$ is the class number. The training data set $D$ expresses as two subsets: $D = D^{l} \cup D^{u}$, where $D^{l}=\{(x_{i}^{l},y_{i}^{l})\}_{i=1}^{N}$ and $D^{u}=\{x_{i}^{u}\}_{i=1}^{M}$ are labeled dataset and unlabeled dataset, respectively. The corresponding gaze dataset $D_{g}$ are collected using eye tracker. We randomly pick a pair of images $(x_{b}^{u},x_{f}^{u})$ as the background image and foreground image, a labeled image $x_{f}^{l}$ as another foreground image, and corresponding gaze data $(g_{f}^{l},g_{f}^{u},g_{b}^{u})$ from the training set. Firstly, according to the position information provided by the gaze data, a one-value rectangular mask $\mathcal{M} \in\{0,1\}^{W\times H}$ that contains all the gaze points is determined, and the human gaze crop is trimmed out according to the rectangular mask. Then, we copy-paste a human gaze crop form $x_{f}^{l}$ (the foreground) onto $x_{b}^{u}$ (the background) to generate the mixed image $x_{mix}^{l}$, and from $x_{f}^{u}$ (the foreground) onto $x_{b}^{u}$ (the background) to generate another mixed image $x_{mix}^{u}$. The foreground human gaze crop is pasted to the background human gaze crop's original position by resizing it to the background crop size. The GazeMix labeled and unlabeled images are as follows:
\vspace{-0.04in}
\begin{equation}
\small
    x_{mix}^{l}=M(R(x_{f}^{l}\,\odot \, \mathcal{M}_{f}^{l}) ,x_{b}^{u}\, \odot \, (1-\mathcal{M}_{b}^{u}))
\end{equation}
\begin{equation}
\small
    x_{mix}^{u}=M(R(x_{f}^{u}\,\odot \, \mathcal{M}_{f}^{u}) ,x_{b}^{u}\, \odot \, (1-\mathcal{M}_{b}^{u}))
\end{equation}
where $x_{f}^{l} \in D^{l}$, $x_{f}^{u},x_{b}^{u} \in D^{u}$, $ \mathcal{M}_{f}^{l},\mathcal{M}_{f}^{u},\mathcal{M}_{b}^{u} \in \mathcal{M}$, and $\odot$ is element-wise multiplication, $R(\cdot)$ means resize operation, $M(\cdot)$ places the resized foreground into the background image.

Unlabeled images are able to increase the data diversity from both $x_{f}^{u}$ and $x_{f}^{l}$. Images $x_{mix}^{u}$ and $x_{mix}^{l}$ are then fed into student network to predict segmentation results $y_{mix}^{u}$ and $y_{mix}^{l}$,
supervised by the pseudo-label $y_{pse}^{u}$ generated from the teacher network and the ground truth $y_{gt}^{l}$ from labeled dataset.

\subsection{Multi-scale Gaze Perception Module}
The MGP module aims to extract multi-scale network perception {of medical images}, which is able to perceive various tissues at fine-grained and holistic levels. 
It thus can flexibly and rapidly facilitate the network in aligning and adapting to {human perception with further guidance from the human gaze}, and precisely perceive critical region, working as a gaze teacher. 

Specifically, we employ channel attention and parallel spatial attention to extract and refine the network perception in multi-scale, ensuring that the perception from different scales is effectively integrated. 
In the channel attention procedure, the feature map $feat$ in front of the segmentation head is firstly executed by global average pooling
(GAP) operation to compress the feature scale. Subsequently, the compressed feature map undergoes two fully connected operations, ReLU activation and sigmoid activation, obtaining the channel attention map $\alpha$. The procedure can be formulated as:
\begin{equation}
\small
    \alpha=\sigma_{sig}(\mathrm{W}_{\!\! fc2}(\sigma_{ relu}(\mathrm{W}_{\!\!fc1}(\text{GAP}(feat))))
\end{equation}
where $\mathrm{W}_{\!\!fc1}$, $\mathrm{W}_{\!\!fc2}$, $\sigma_{relu}$ and $\sigma_{sig}$ represent the fully connected layers, Relu and Sigmoid activation, respectively. 

The value in $\alpha$ emphasizes the importance of the corresponding channel information in the segmentation feature $feat$. The calibration activity is executed on $feat$ with $\alpha$, as:
\begin{equation}
\small
    feat_{c}=\alpha \otimes feat
\end{equation}
where $feat_{c}$ denotes channel optimized feature, $\otimes$ means channel-wise multiplication.

Then, in parallel spatial attention procedure, $feat_{c}$ is executed
with $1\times1$, $3\times3$, $7\times7$ spatial convolution operation and ReLU activation. Sigmoid activity is further performed to calculate the spatial attention $\beta_{1}$, $\beta_{3}$, $\beta_{7}$ at different scales, as:
\begin{equation}
\small
\beta_{i}=\sigma_{sig}(\sigma_{relu}(\mathrm{W}_{\! i\times i}(\text{GAP}(feat_{c}))+b_{i\times i}))
\end{equation} 
where $\beta_{i}$, $W_{i\times i}$, $b_{i}$ mean different-scale spatial attention coefficients, convolution kernel and bias terms, with $i=1,3,7$.
Finally, 
the multi-scale network perception $g_{net}$ is implemented through fusing $\beta_{1}$, $\beta_{3}$, and $\beta_{7}$ of different scale, as: 
\begin{equation}
\small
    g_{net}=\sigma_{sig}(\mathrm{W}_{\! 1\times1}(\beta_{1}\oplus  \beta_{3} \oplus \beta_{7} ))
\end{equation}
where $\oplus$ means channel concatenation.

\vspace{-0.1in}
\subsection{Loss Function}
\vspace{-0.04in}

Model parameters are comprehensively updated under the guided by two ‘teachers’, with two heterogeneous loss functions in which the Gaze Loss $\mathcal{L}_{gaze}$ works to teach “where should be gazed'' and the Segmentation Loss $\mathcal{L}_{seg}$ deploys to teach ``what content gazed on''.

Gaze Loss $\mathcal{L}_{gaze}$ manages to align network perception $g_{net}$ with human gaze $g_{human}$, so that it teaches the model to understand “where", i.e., the critical foreground range that radiologists clinically care about medical images, like human. 
By comparing $g_{gaze}$ and $g_{net}$ through $\mathcal{L}_{gaze}$, 
the cognitive difference in foreground perception is identified and rectified, such as
the important foreground regions that the model does not focus on but the radiologist cares about, and the interfering background areas that the model misinterprets but radiologist distinguishes.
$\mathcal{L}_{gaze}$ thus {guides the model to learn }``where'' to remedy {the misunderstanding in distribution of medical images caused by background interference}, and correct the errors in identifying important regions of medical interpretation for foreground deeply focusing. It is calculated as:
\begin{equation}
\small
    \mathcal{L}_{gaze}=\text{MSE}(g_{net},g_{human})
\end{equation}
where MSE represents mean squared error.

Segmentation Loss $\mathcal{L}_{seg}$ utilizes GazeMix pre-trained teacher network to guide the model to learn content including semantics, shape, boundaries, etc., for unlabeled data, {so that it teaches the model to understand} detailed and specific ``what'' is important in medical images, and enhances the model robustness {eliminating the homogeneity caused by only a few labels}. Pre-trained with GazeMix, the teacher network {is deeply empowered with rich knowledge on foreground target}, as GazeMix robustly enables the essentialized and reliable augmentation on the limited labeled dataset with human perception injection from gaze. 
It is thus good at providing pseudo-labels for unlabeled data.
As it is built in mean-teacher network architecture, $\mathcal{L}_{seg}$ also synchronously employs ground truth from fewer labeled data to supervise model learning, together with GazeMix to enrich these data.  
Specifically, in each training iteration, samples are randomly selected from the limited labeled dataset and massive unlabeled data. Labeled data $x_f^l$ and unlabeled data $x_f^u$  , $x_b^u$ together construct the richly mixed image $x_{mix}^l$ and $x_{mix}^u$, through GazeMix with human gaze. The pseudo-label $y_{pse}^u$ generated by the teacher network and ground truth $y_{gt}^l$ from labeled dataset are employed for $x_{mix}^u$ and $x_{mix}^l$, respectively. Segmentation Loss thus consists of two parts $\mathcal{L}_{pse}$ and $\mathcal{L}_{gt}$ for $x_{mix}^u$ and $x_{mix}^l$, as:   
\begin{equation}
\small
\begin{split}
    % \mathcal{L}_{seg}=(Dice(P_{0,1}^{mix},Y_{pse,gt}^{mix})+\\
    % CE(P_{0,1}^{mix},Y_{pse,gt}^{mix}))/2,
    \mathcal{L}_{seg}=(\mathcal{L}_{gt}(y_{mix}^{l},y_{gt}^{l})+\mathcal{L}_{pse}(y_{mix}^{u},y_{pse}^{u}))/2
\end{split}
\end{equation}
where $\mathcal{L}_{pse}$ and $\mathcal{L}_{gt}$ are computed by the combination of Dice loss and Cross-entropy loss.

Together at each iteration, parameters $\Theta_{s}$ in the student network of model are updated by stochastic gradient descent with the collaboration of $\mathcal{L}_{gaze}$ and $\mathcal{L}_{seg}$: 
\begin{equation}
\small
\mathcal{L}_{all}=\mathcal{L}_{seg}+\lambda\mathcal{L}_{gaze}
\label{loss_all}
\end{equation}

\begin{algorithm}[t]
%\vspace{-0.3in}
\small
  \caption{Training Procedure}
  \label{alg}
  \begin{algorithmic}[0]
    \Require
      Image Dataset $D = D^{l} \cup D^{u}$, Gaze data $D^{g}$
    
    \State $x_{f}^{l}, y_{gt}^{l} \in D^{l}$,\; $x_{f}^{u},x_{b}^{u} \in D^{u}$,\;
    $g_{f}^{l},g_{f}^{u},g_{b}^{u} \in D^{g}$
    \\
    $\bullet $ \textbf{GazeMix generates a mixture of images}: 
    \State $x^{l}_{mix}=$ GazeMix$(x_{f}^{l},x_{b}^{u}, g_{f}^{l},g_{b}^{u})$
    \State $x^{u}_{mix}=$ GazeMix$(x_{f}^{u},x_{b}^{u}, g_{f}^{u},g_{b}^{u})$
    \\
    $\bullet $ \textbf{The mixed images are fed into the student network}: 
    \State $y_{mix}^{l}, g_{net}^{l} = \mathcal{S}(x_{mix}^{l};\;\Theta _{s})$
    \State $y_{mix}^{u}, g_{net}^{u} = \mathcal{S}(x_{mix}^{u};\;\Theta _{s})$
    \\
    $\bullet $ \textbf{The teacher network generates pseudo labels}: 
    \State $y_{pse}^{u} = \mathcal{T}(x_{mix}^{u};\;\Theta _{t})$
    \\
    $\bullet $ \textbf{Loss calculation with Eq. (\ref{loss_all})}: 
    \State $\mathcal{L}_{all}=\mathcal{L}_{seg}+\lambda\mathcal{L}_{gaze}$
    \State
    $\bullet $ \textbf{Gradient descent update $\mathcal{S}(\Theta _{s})$}
    \State
    $\mathcal{S}(\Theta _{s}) = \mathcal{S}(\Theta _{s}) - \eta \nabla \mathcal{L}_{all}(\Theta _{s})$
    \State
    $\bullet $ \textbf{EMA update $\mathcal{T}(\Theta _{t})$}
    \State
    $\mathcal{T}(\Theta _{t} = \gamma \mathcal{T}(\Theta _{t}) + (1-\gamma) \mathcal{S}(\Theta _{s})$
    
  \end{algorithmic}
\end{algorithm}

\begin{table}[tb]
\setlength{\tabcolsep}{2.5pt}
\centering
%\small
\caption{Comparisons with state-of-the-art semi-supervised segmentation methods on the ACDC, CAMUS, Synapse, and SCR datasets.}\label{result}
%\vspace{-0.05in}
\resizebox{\textwidth}{!}{
\begin{tabular}{c|c| c c c c| c c c c| c c c c| c c c c}
\hline
\multirow{2}*{Method} & \multirow{2}{*}{\begin{tabular}[c]{@{}c@{}}Labeling\\ Ratio\end{tabular}} & \multicolumn{4}{c|}{\textbf{ACDC} (MRI)} & \multicolumn{4}{c|}{\textbf{CAMUS} (Ultrasound)} & \multicolumn{4}{c|}{\textbf{Synapse} (CT)} &\multicolumn{4}{c}{\textbf{SCR} (X-Ray)} \\ 
 
\cline{3-18} ~ & & Dice$\uparrow$ &Jaccard$\uparrow$  & 95HD$\downarrow$ & ASD$\downarrow$ & Dice$\uparrow$ &Jaccard$\uparrow$  & 95HD$\downarrow$ & ASD$\downarrow$  & Dice$\uparrow$ &Jaccard$\uparrow$  & 95HD$\downarrow$ & ASD$\downarrow$ & Dice$\uparrow$ &Jaccard$\uparrow$  & 95HD$\downarrow$ & ASD$\downarrow$ \\
\hline

UA-MT & \multirow{6}*{5\%} & 46.04 &35.97 &20.08 &7.75 & 78.63 &65.60 &21.74 &16.58 & 65.76 & 61.84 & 28.78 & 30.39 & 69.27 & 65.73 & 40.28 & 35.80 \\

DTC & & 56.90 &45.67 & 23.36 & 7.39 & 80.81 &67.37 & 17.32 & 14.42 & 67.93 & 63.70 & 27.24 & 29.22 & 74.34 & 61.46 & 38.46 & 36.27\\

SS-Net & &65.83 &55.38 &6.67 &2.28 &85.89 &76.56 &13.59 &11.85 & 77.95 & 69.25 & 24.38 & 25.61 & 84.79 & 74.85 & 22.81 & 29.19\\

MC-Net &  &62.85 &52.29  &7.62 & 2.33 &82.84 &71.36  &15.51 & 13.85 &76.99 & 67.76 & 26.41 & 28.14 & 84.53 & 70.29 & 27.43 & 32.66\\

BCP & &87.59 &78.67 &1.90 &0.67 &86.15 &76.66 &10.22 &10.79 & 79.61 & 69.17 & 23.08 & 23.73 & 86.33 & 76.19 & 20.78 & 25.75\\

\textbf{Ours} & & \textbf{89.37} &\textbf{81.26} &\textbf{1.45} &\textbf{0.39} & \textbf{86.41} & \textbf{77.19} & \textbf{9.27} &\textbf{9.87} & \textbf{80.72} &\textbf{70.76} &\textbf{21.22} &\textbf{23.49} & \textbf{88.31} & \textbf{77.36} & \textbf{16.25} & \textbf{22.29}\\
\hline

UA-MT & \multirow{6}*{10\%} & 81.65 &70.64 &6.88 &2.02 & 83.25 &76.21 &11.64 &13.83 & 78.75 & 62.73 & 22.04 & 24.67 & 81.11 & 68.26 & 36.74 & 33.68\\

DTC & & 84.29 &73.92 & 12.81 & 4.01 & 84.63 &77.35 & 10.21 & 12.47 & 75.96 & 65.02 & 21.19 & \textbf{20.25} & 82.74 & 65.08 & 34.33 & 36.24\\

SS-Net & &86.78 &77.67 &6.07 &1.40 &86.98 &78.31 &9.71 &13.48 & 80.87 & 69.59 & 21.96 & 23.11 & 87.06 & 76.35 & 20.48 & 26.73\\

MC-Net & &86.44 &77.04  &5.50 & 1.84 &86.65 &76.85  &9.28 & 13.19 & 79.24 & 69.79 & 25.02 & 26.61 & 86.61 & 72.30 & 23.62 & 30.94\\

BCP & & 88.84 &80.62 &3.98 &1.17 & 87.88 &79.19 &8.40 &11.53 & 82.68 & 71.09 & 20.87 & 21.46 & 88.18 & 77.42 & 18.13 & 21.53\\

\textbf{Ours} & &\textbf{90.87} &\textbf{83.67} &\textbf{2.07} &\textbf{0.69} &\textbf{88.05} &\textbf{79.51} &\textbf{8.22} &\textbf{7.32} &\textbf{83.56} & \textbf{72.41} & \textbf{19.39} & 20.78 & \textbf{90.75} & \textbf{78.76} & \textbf{15.44} & \textbf{20.41}\\
\hline
\end{tabular}
}
%\vspace{-0.2in}
\end{table}

\subsection{Network Training}
As shown in Alg.\ref{alg}, our training procedure is conducted as: 
(1) The labeled data and unlabeled data are randomly sampled and passed through GazeMix to generate mixed data.
(2) Both labeled and unlabeled mixed data are fed into the student network $\mathcal{S}(\Theta _{s})$ to obtain the segmentation results and network perceptions. 
(3) Synchronously, the pre-tranied teacher network $\mathcal{T}(\Theta _{t})$ generates pseudo labels for unlabeled mixed data. 
(4) Training loss $\mathcal{L}_{all}$ composed by $\mathcal{L}_{seg}$ and $\mathcal{L}_{gaze}$  is calculated with Eq. (\ref{loss_all}).
(5) The student network $\mathcal{S}(\Theta _{s})$ is optimized by gradient descent. 
And (6) the teacher network is updated by EMA of student network.

\section{Experiments}
\label{sec:Experiments}

\subsection{ Experiments Setting}
Public datasets including ACDC \cite{bernard2018deep}, CAMUS \cite{leclerc2019deep}, Synapse \cite{landman2015miccai} and SCR \cite{shiraishi2000development} are used  in our experiment. The corresponding gaze data are collected by two radiologists. 
ACDC dataset contains segmentation of right ventricle, left ventricle, and myocardium, from 100 patients’ scans. 
We follow the setting in BCP\cite{bai2023bidirectional}, as 70, 10, and 20 patients’ scans for training, validation, and testing.   
CAMUS dataset consists of clinical ultrasound images from 500 patients. It contains segmentation of left ventricle, left atrium, and endocardium. It is split into 400, 50, and 50 patients’ scans for training, validation, and testing.
{Synapse dataset is composed of 3779 axial abdominal clinical CT images with a size of 512×512 pixels and contains three organs (liver, kidney, stomach). 2645 images are used for training, while 369 and 756 images are used for validation and testing respectively. 
SCR (Segmentation in Chest Radiographs) is a dataset for the study of segmentation of chest anatomical structures. The dataset contains posterior-anterior chest radiographs of 347 subjects with manually annotated lungs and hearts. We split the dataset into three parts as of 100, 47, and 100, which are used for training, validation, and testing respectively. 
}

{We recruited four experienced practicing radiologists for the experiment. Each radiologist was assigned specific tasks based on their expertise and the types of medical images involved. Junior radiologist A (Dr. Yehui Jiang), with five years of clinical experience, was in charge of reading ultrasound images (CAMUS dataset) during the data-collection process. Junior radiologist B (Dr. Fangyi Xu), having five years of clinical experience, focused on gaze data collection from MRI images (ACDC dataset). And Junior radiologist C (Dr. Cong Xia), with seven years’ experience, was responsible for collecting gaze data from CT and X-Ray images (Synapse and SCR datasets). Senior radiologist (Dr. Yinsu Zhu), with over 15 years of clinical experience, played a crucial role in quality control.
After junior radiologists A, B and C completed their data-collection tasks, senior radiologist rechecked the collected gaze data to ensure its validity. For cases that did not meet the pre-set criteria, such as incorrect gazing areas, insufficient gaze points, and excessive noise, senior radiologist removed them, and junior radiologists recollected the data. }

All experiments are implemented by PyTorch on an NVIDIA 3090 GPU.
We use 2D U-Net as backbone of our model. 
The batch size is set to 24. 
The training iteration is 30k.
$\lambda$=0.5 is set to balance the loss. 
Dice coefficient, Jaccard Score, $95\%$ Hausdorff Distance (95HD), and Average Surface Distance (ASD) are used to evaluate segmentation performance.

\subsection{Overall Performance}
Our model achieved advanced performance on both datasets at different labeling ratio, Specifically: 
1) For ACDC dataset, it achieves high Dice of $89.37\%$ \&  $90.87\%$, high Jaccard of $81.26$ \& $83.67$, low 95HD of $1.45$ \& $2.07$, and low ASD of $0.39$ \& $0.69$, at ultra low labeling ratio $5\%$  and $10\%$. 
2) For CAMUS dataset, it also shows remarkable superiority on all four metric as $86.41\%$ \& $88.05\%$ (Dice), $77.19$ \& $79.51$ (Jaccard), $9.27$ \& $8.22$ (95HD), and $9.87$ \& $7.32$ (ASD), at ultra low labeling ratio 5\% and 10\%.    
{3) For Synapse dataset, at the labeling ratio of 5\% and 10\%, our model attains high Dice of 80.72\% \& 83.56\%, high Jaccard of 70.76 \& 72.41, low 95HD of 21.22 \& 19.39, and low ASD of 23.49 \& 20.78, respectively. 
4) For SCR dataset, with labeling ratio of 5\% and 10\%, our model accomplishes high Dice of 88.31\% \& 90.75\%, high Jaccard of 77.36 \& 78.76, low 95HD of 16.25 \& 15.44, and low ASD of 22.29 \& 20.41. These remarkable outcomes emphasize that our model can effectively handle different modality data and organs, and sustain excellent segmentation performance despite the limited quantity of labeled data.
}

\begin{table*}[t]
\setlength{\tabcolsep}{2pt}
\centering
%\small
\caption{Ablation study for the effectiveness of each component. The baseline network is trained using the basic mean-teacher model.}\label{abs_mod}
%    \vspace{-0.05in}
\resizebox{\textwidth}{!}{
\begin{tabular}{c|c| c c c | c c c c| c c c c }
\hline
\multirow{2}*{Method}& \multirow{2}{*}{\begin{tabular}[c]{@{}c@{}}Labeling\\ Ratio\end{tabular}}  & \multicolumn{3}{c|}{\multirow{1} *{Component}} & \multicolumn{4}{c|}{ACDC} & \multicolumn{4}{c}{CAMUS} \\ 
\cline{6-13}
\cline{3-13} ~ &  & GazeMix & MGP  & $L_{gaze}$  & Dice$\uparrow$ & Jaccard$\uparrow$ & 95HD$\downarrow$ & ASD$\downarrow$ & Dice$\uparrow$ &Jaccard$\uparrow$  & 95HD$\downarrow$ & ASD$\downarrow$ \\
\hline
Baseline & & $\times$ & $\times$ & $\times$ & 70.34 &60.32 &10.28 &8.13 & 71.53 &61.74 &19.82 &16.37\\
+GazeMix & & $\checkmark$ & $\times$ & $\times$ & 88.27 &79.63 &1.73 &0.85 & 85.34 &74.63 &11.21 &13.29\\
+MGP & & $\times$ & $\checkmark$ & $\times$ & 86.52 &75.91 &7.12 &2.74 & 82.29 &70.55 &14.26 &14.95\\
+MGP+$L_{gaze}$ & 5$\%$ & $\times$ & $\checkmark$ & $\checkmark$ &  88.16 &76.65 &5.91 &2.19 & 83.08 &71.95 & 12.74 &13.28\\
+GazeMix+MGP &  & $\checkmark$ & $\checkmark$ & $\times$ &  88.63 &80.48 &1.51 &0.42 & 85.93 &74.79 & 10.39 &11.64\\
\textbf{+GazeMix+MGP+$\mathcal{L}_{gaze}$ } & & $\checkmark$ & $\checkmark$ & $\checkmark$ & \textbf{89.37} & \textbf{81.26} & \textbf{1.45} & \textbf{0.39} & \textbf{86.41} &\textbf{77.19} & \textbf{9.27} &\textbf{9.87}\\

\hline
Baseline & & $\times$ & $\times$ & $\times$ & 81.72 &69.27 &9.28 &3.71 &  79.64 &70.26 &14.26 &13.83\\
+GazeMix & & $\checkmark$ & $\times$ & $\times$ & 89.33 &80.48 &3.27 &1.62 &  86.34 &72.79 & 12.43 &11.57\\
+MGP & & $\times$ & $\checkmark$ & $\times$ & 88.14 &78.61 &5.22 &2.08 & 85.71 &72.04 &13.76 &12.83\\
+MGP+$L_{gaze}$ & 10$\%$ & $\times$ & $\checkmark$ & $\checkmark$ &  88.73 &80.10 &3.94 &1.84 & 86.83 &74.61 & 12.81 &11.86\\
+GazeMix+MGP &  & $\checkmark$ & $\checkmark$ & $\times$ & 89.85 &81.23 &2.31 &1.07 &  87.42 &78.26 & 10.12 &9.74\\
\textbf{+GazeMix+MGP+$\mathcal{L}_{gaze}$ } & & $\checkmark$ & $\checkmark$ & $\checkmark$ & \textbf{90.87} & \textbf{83.67} & \textbf{2.07} & \textbf{0.69} & \textbf{88.05} &\textbf{79.51} &\textbf{8.22} &\textbf{7.32}\\
\hline
\end{tabular}
}
%\vspace{-0.13in}
\end{table*} 

\subsection{Comparison with Sate-of-the-Art Methods}

\textbf{a) Quantitative Comparison:}
We make comparison between our model and various SoTA competitors including UA-MT \cite{yu2018pu}, 
DTC \cite{luo2021semi}, 
MC-Net \cite{wu2022exploring}, SS-Net \cite{wu2021semi}, and BCP \cite{bai2023bidirectional}.
The semi-supervised segmentation is evaluated with 5\% and 10\% labeling ratio on public datasets of ACDC, CAMUS, Synapse, and SCR. 

As the results on ACDC dataset shown in Table \ref{result}, our model surpasses all state-of the-art methods, by increasing Dice with $25.53\%$ \& $5.27\%$, increasing Jaccard with $27.67$ \& $7.70$, decreasing 95HD with $10.48$ \& $4.98$, and decreasing ASD with $3.70$ \& $1.40$, at labeling ratio $5\%$ and $10\%$ respectively.  
For comparison on CAMUS dataset, our model gains the best performance, increasing Dice with $ 3.55\%$ \& $ 2.63\%$, increasing Jaccard with $5.68$ \& $1.93$, decreasing 95HD with $6.41$ \& $1.63$, and decreasing ASD with $3.63$ \& $5.58$, at labeling ratio $5\%$ and $10\%$ respectively. 
{As shown in Table \ref{result}, our model still achieves the best performance across all metrics on the Synapse and SCR datasets as well. 
On the Synapse dataset, our model increases the Dice by 7.07\% \& 4.06\%, the Jaccard by 4.42 \& 4.76, and decreases the 95HD by 4.76 \& 2.83, the ASD by 3.93 \& 2.44 at the labeling ratio of 5\% and 10\%, respectively. 
Our model also achieves excellent performance on the SCR dataset. At the labeling ratio of 5\% and 10\% respectively, our model increases the Dice by 8.46\% \& 5.61\%, the Jaccard by 7.66 \& 6.88, and decreases the 95HD  by 13.70 \& 11.22, the ASD by 9.64 \& 9.41. 
}

{Our approach shows strong performance across different organs, further proving its versatility and robustness in handling diverse anatomical structures. This result demonstrates that the proposed HG-DTGL is able to effectively utilize unlabeled data in the presence of limited labeled data to achieve more accurate segmentation performance. In terms of multimodal data adaptability and cross-organ versatility, the proposed method has achieved excellent performance in various medical imaging modalities such as MRI, ultrasound, CT and X-ray, and different organ segmentation tasks such as liver, kidney, heart, and so on. This comprehensive experimental analysis based on different evaluation metrics and diverse scenarios fully proves significant advantages of our method. These advantages are mainly attributed to our innovative processing of gaze data: through the MGP module and Gaze Loss, the model can effectively capture the critical areas, and GazeMix can effectively and steadily expand the size of the dataset.}

\begin{figure}[p]
    \centering
    \includegraphics[width=1\linewidth]{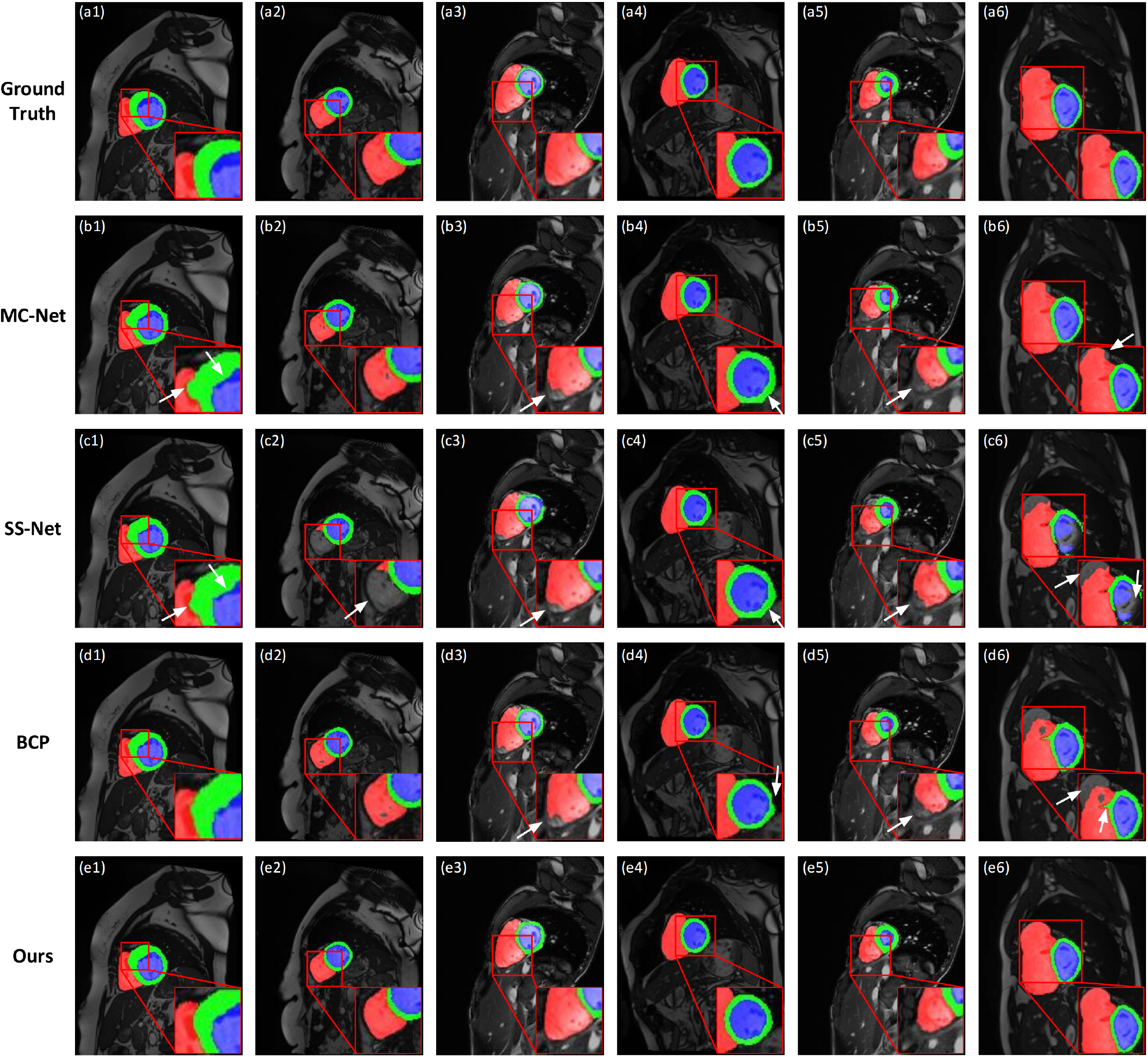}
    \caption{Visualization comparison between our proposed HG-DTGL model and different semi-supervised segmentation methods, on the ACDC dataset using 5\% labeled data. HG-DTGL behaves precisely in details and semantic boundary among different tissues, consistent to the ground truth.}
\label{visual}
\end{figure}

\begin{figure}[t]
    \centering
    \includegraphics[width=1\linewidth]{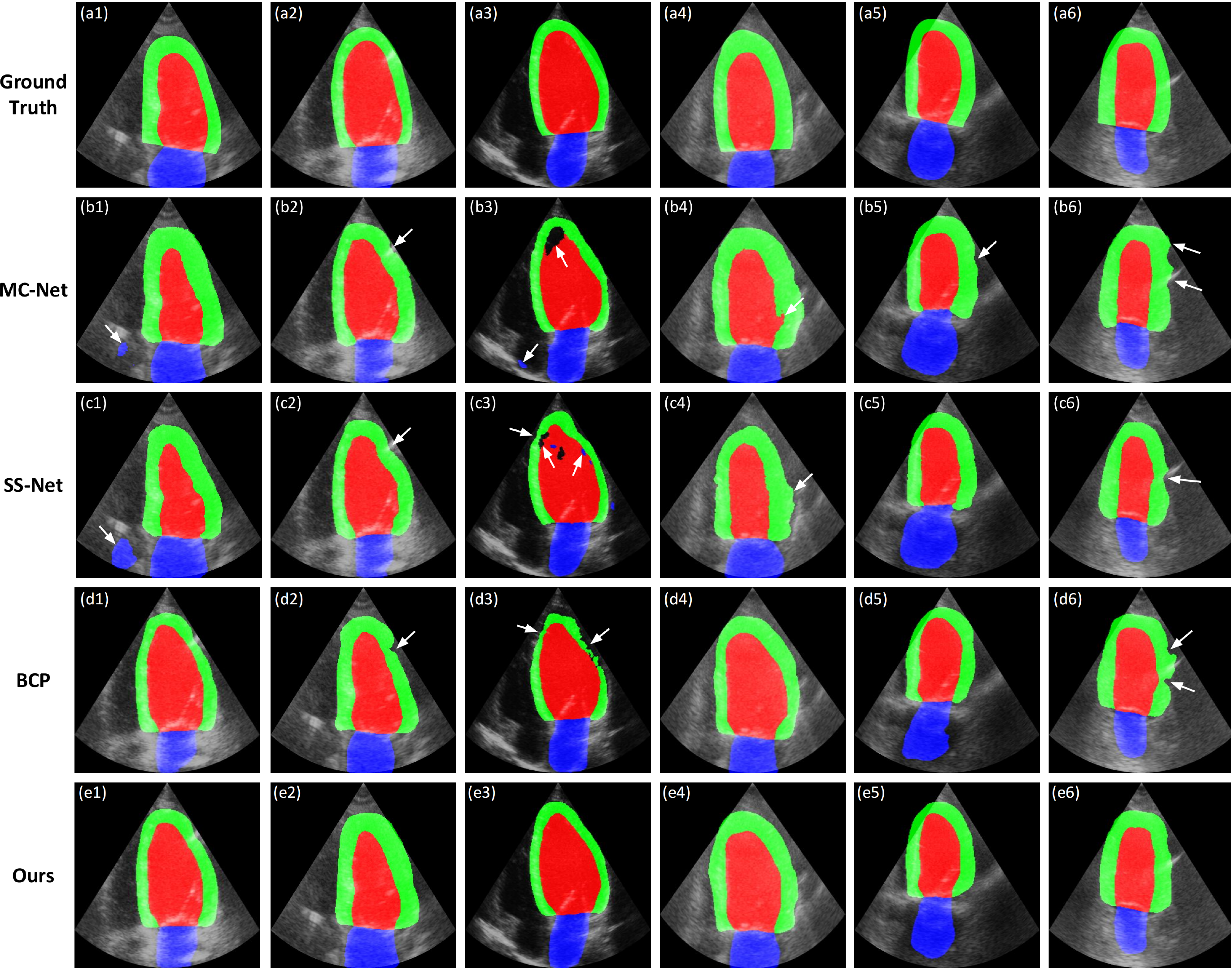}
    \caption{Visualization comparison between our proposed HG-DTGL model and different semi-supervised segmentation methods, on the CAMUS dataset using 5\% labeled data. 
    HG-DTGL behaves the best performance with  segmentation completeness, boundaries, details and distribution, consistent to the ground truth.
    }
\label{camus_visual}
\end{figure}

\begin{figure}[t]
    \centering
    \includegraphics[width=0.5\linewidth]{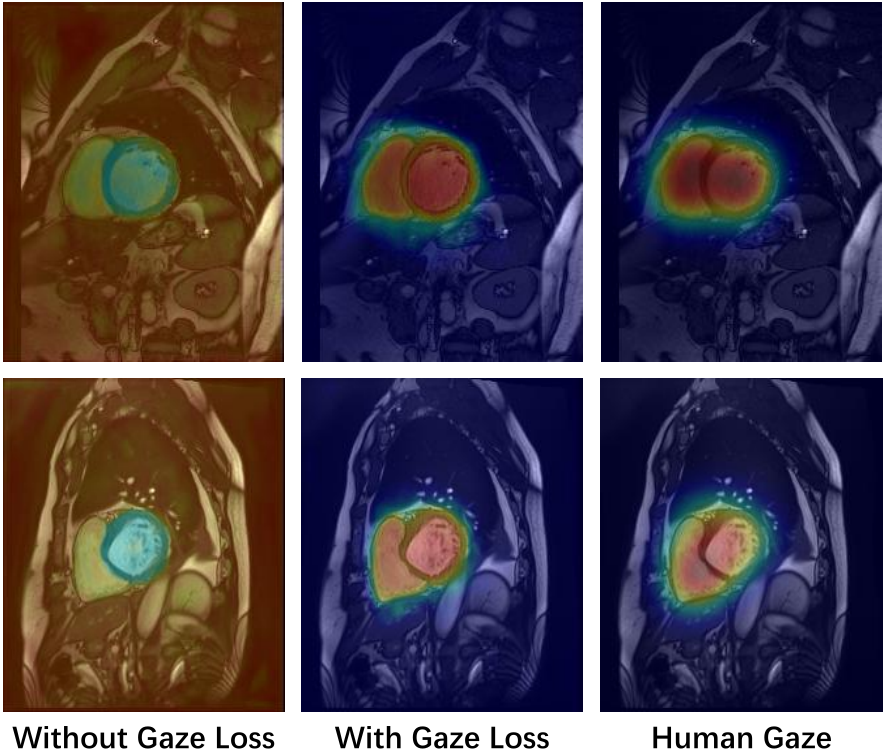}
    \caption{Gaze Loss guides the mode to deeply focus on the target tissue like human perceives, and eliminate inference of the irrelevant regions.}
\label{perception}
\end{figure}

\begin{figure*}[t]
    \centering
    \includegraphics[width=1\linewidth]{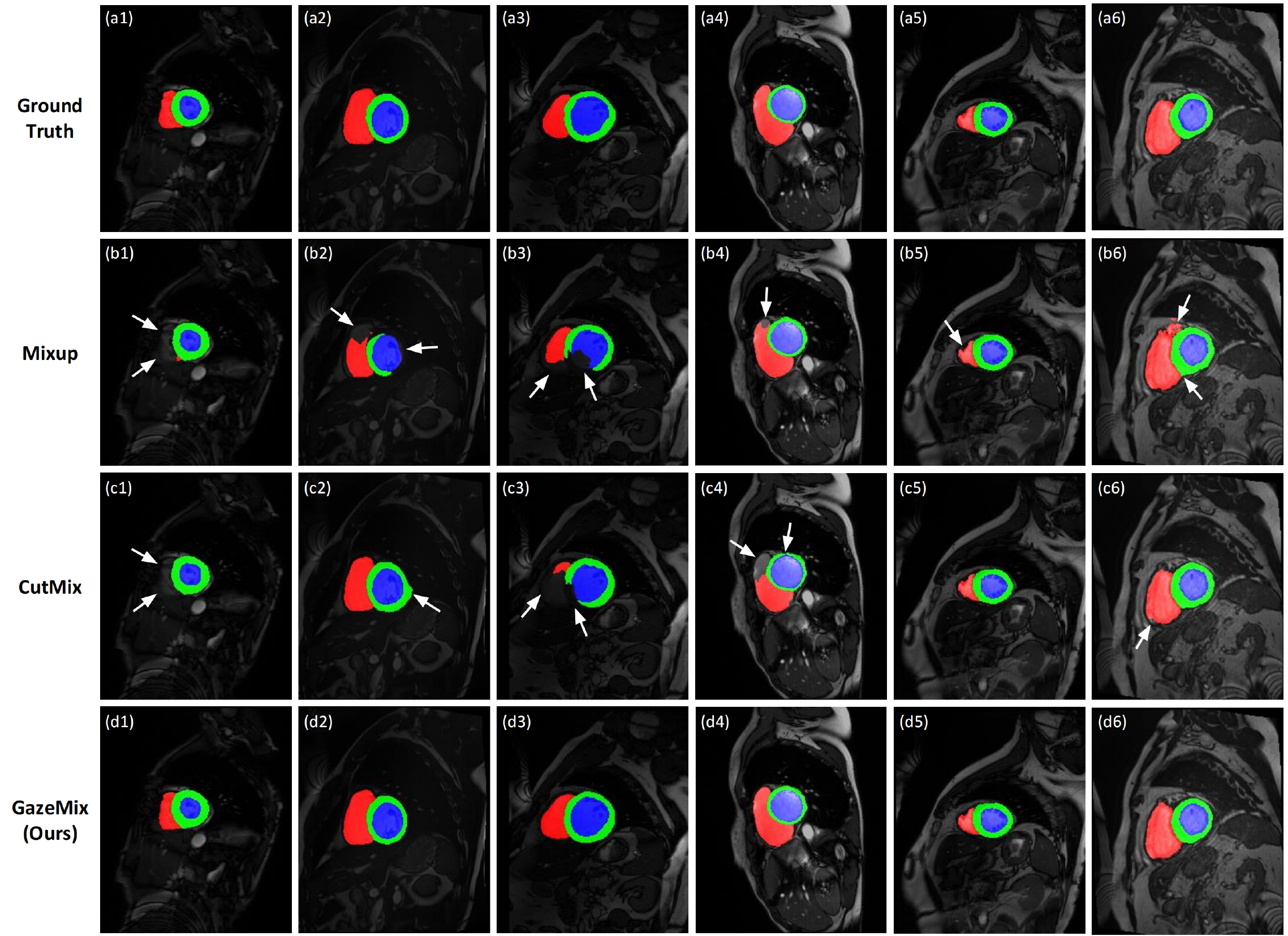}
    \vspace{-0.12in}
    \caption{
    Visualization comparison of our proposed GazeMix with Mixup and CutMix, on the ACDC dataset using 5\% labeled data. 
The GazeMix behaves the best visual segmentation effect with accurate structure and detail, consistent to the ground truth.}
\label{mix_type}
\vspace{-0.05in}
\end{figure*}

\begin{table}
\setlength{\tabcolsep}{2.5pt}
\centering
\small
\caption{ Comparison of GazeMix with GCA, Mixup, and CutMix.}\label{abs_mix}
%\vspace{-0.05in}
\resizebox{\textwidth}{!}{
\begin{tabular}{c|c| c c c c| c c c c}
\hline
\multirow{2}*{Method} &  \multirow{2}{*}{\begin{tabular}[c]{@{}c@{}}Labeling\\ Ratio\end{tabular}} & \multicolumn{4}{c|}{ACDC} & \multicolumn{4}{c}{CAMUS} \\ 
 
\cline{3-10} ~ &   & Dice$\uparrow$ & Jaccard$\uparrow$ & 95HD$\downarrow$ & ASD$\downarrow$ & Dice$\uparrow$ & Jaccard$\uparrow$ & 95HD$\downarrow$ & ASD$\downarrow$ \\
\hline
Mixup &\multirow{4}*{5\%} &86.37 &79.06 &9.23 &1.91 &85.12 &75.59 &13.25 &12.38\\
CutMix &  &87.81 & 80.32 &5.38 &1.24 &85.31 &75.62 &12.74 &10.92\\
GCA & &86.79 &77.29 & 6.58 &2.15 &84.37 &73.84 &15.63 &13.82\\
\textbf{GazeMix} & &  \textbf{89.37} &\textbf{81.26} &\textbf{1.45} &\textbf{0.39} &\textbf{86.41} &\textbf{77.19} &\textbf{9.27} &\textbf{9.87}\\
\hline
Mixup & \multirow{4}*{10\%} &87.29 &81.17 &6.18 &1.62 &86.12 &77.82 &12.93 &8.90\\
CutMix &  &88.93 &82.46 &3.91 &0.93 &86.55 &77.79 &11.61 &8.65\\
GCA &  &88.62 &81.35 &5.49 &1.07 &85.93 &76.55 &13.09 &10.43\\
\textbf{GazeMix} & &  \textbf{90.87} & \textbf{83.67} & \textbf{2.07} & \textbf{0.69} & \textbf{88.05} & \textbf{79.51} & \textbf{8.22} & \textbf{7.32}\\
\hline
\end{tabular}
}
\vspace{-0.2in}
\end{table}

\textbf{b) Visualization Comparison:}
As shown in Figure \ref{visual} and \ref{camus_visual}, our model achieves the best visual segmentation consistent with the ground truth, compared with the SoTA competitors MC-Net \cite{wu2022exploring}, SS-Net \cite{wu2021semi}, and BCP \cite{bai2023bidirectional} (the Top3 comparison methods in quantitative evaluation).

\begin{table}[b]
\vspace{-0.3in}
\setlength{\tabcolsep}{2.5pt}
\centering
\small
\caption{Comparison of MGP with Spatial Attention and CBAM.}\label{abs_attention}

\resizebox{\textwidth}{!}{
\begin{tabular}{c|c| c c c c| c c c c}
\hline
\multirow{2}*{Method} & \multirow{2}{*}{\begin{tabular}[c]{@{}c@{}}Labeling\\ Ratio\end{tabular}} & \multicolumn{4}{c|}{ACDC} & \multicolumn{4}{c}{CAMUS} \\ 
 
\cline{3-10} ~ &   & Dice$\uparrow$ & Jaccard$\uparrow$ & 95HD$\downarrow$ & ASD$\downarrow$ & Dice$\uparrow$ & Jaccard$\uparrow$ & 95HD$\downarrow$ & ASD$\downarrow$ \\
\hline
Spatial Attention & &88.21 &78.46 &1.93 &0.92  &84.91 &74.29  &12.77 &11.68 \\
CBAM & 5\% &88.74 &80.05 &1.82 &1.16 &85.29 &75.14 &11.29 &11.30\\
\textbf{MGP} & & \textbf{89.37} &\textbf{81.26} &\textbf{1.45} &\textbf{0.39} &\textbf{86.41} &\textbf{77.19} &\textbf{9.27} &\textbf{9.87}\\

\hline
Spatial Attention & &88.38 &80.81 &4.07 &0.85 &85.87 &75.80 &11.24 &10.76\\
CBAM & 10\% &89.05 &81.45 &3.64 &1.03 &86.73 &77.95 &10.21 &9.75\\
\textbf{MGP} & &  \textbf{90.87} & \textbf{83.67} & \textbf{2.07} & \textbf{0.69} & \textbf{88.05} & \textbf{79.51} & \textbf{8.22} & \textbf{7.32}\\

\hline
\end{tabular}
}
%\vspace{-0.2in}
\end{table}

On ACDC dataset, our proposed HG-DTGL is able to correctly distinguish the distribution and content of right ventricle (RV), left ventricle (LV), and myocardium, as shown in Figure \ref{visual}(e1)-(e6). {It ensures the structural integrity}, while the incomplete LV and RV, myocardium fracture exist in MC-Net, SS-Net and BCP due to missing segmentation as Figure \ref{visual}.(b)-(d). And our HG-DTGL also behaves precisely in details and semantic boundary among different tissues, while comparison methods confuse the tissues and make too thick/thin and misidentified myocardium as Figure \ref{visual}(b1), (c1), (d4) and (d6).
These significant improvements in our HG-DTGL are beneficial from the GazeMix, MGP, and Gaze Loss. GazeMix enables human perception-based data augmentation, to drive model deeply understand the target content even in only $5\%$ labeled dataset. MGP comprehensively boosts network perception from multi-scale, so that adaptively facilitate model aligned to human gaze. And as shown in Figure \ref{perception}, 
Gaze Loss guides model to deeply focus on the target tissue human perceived, and eliminate interference of the irrelevant regions.

As visual results on the CAMUS dataset in Figure \ref{camus_visual} show, the proposed model also robustly gains the best performance to accurately segment multiple tissues LV, left atrium (LA) and endocardium in ultrasound image, compared with SoTA methods. As Figure \ref{camus_visual} (e) \textit{vs.} (b)-(d), it behaves precise segmentation on shape, details, and semantic boundary recognition, while incomplete/wrong shape, broken structure, and incorrect tissues distribution occur in MC-Net, SS-Net and BCP due to the inherent characteristics of ultrasound image that indirect relation between pixel intensity and the physical property of the tissue. {Such imaging characteristics require segmentation with high understanding of content and distribution}. The introduction of gaze in GazeMix and MGP+Gaze Loss in our HG-DTGL is good at guiding model to deeply understand and learn ``what'' and ``where''. Specifically, as Figure \ref{camus_visual} (e1) \textit{vs.} (b1)\&(c1), the proposed model enables consistent distribution and shape with ground truth, while missegmentation RA region and underestimated LV shape happen in MC-Net and SS-Net. As Figure \ref{camus_visual} (e3)\&(e6) \textit{vs.} (b3)-(d3)\&(b6)-(d6), our model achieves accurate detailed structure, while incomplete LV and broken endocardium exist in the compared methods with arrow indicated.

\begin{figure*}[t]
    \centering
    \includegraphics[width=1\linewidth]{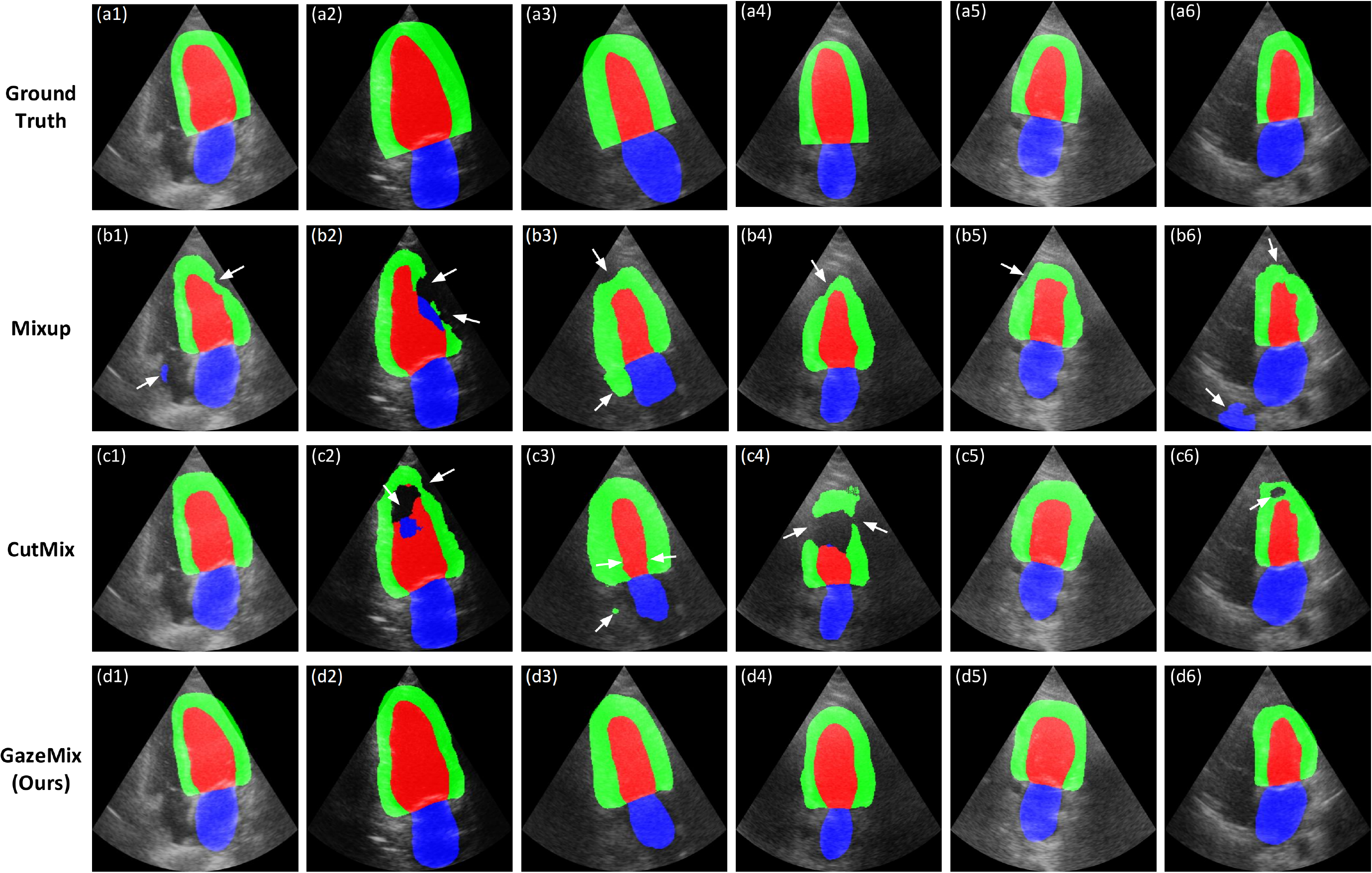}
    \vspace{-0.1in}
    \caption{Visualization comparison of our proposed GazeMix with Mixup and CutMix, on the CAMUS dataset using 5\% labeled data. The GazeMix also behaves the best visual segmentation effect with accurate structure and detail, consistent to the ground truth.}
\label{mix_camus}
\vspace{-0.15in}
\end{figure*}

\vspace{-0.03in}
\subsection{Ablation Studies}
\textbf{a) The effectiveness of each component:} 
We conduct ablation studies to show the effectiveness of each component in our method. 
Table \ref{abs_mod} shows 
results
on ACDC and CAMUS dataset with 5\% and 10\% labeling ratio. 
The first row indicates the mean-teacher baseline model. Compared with the baseline, the proposed model gains robust improvement with Dice increase of 19.03\% \& 14.88\%, Jaccard increase of 20.94 \& 15.45, 95HD decrease of 8.83 \& 10.55, and ASD decrease of 7.74 \& 6.50 on 5\% labeled ACDC and CAMUS, as well as Dice increase of 9.15\% \& 8.41\%, Jaccard increase of 14.40 \& 9.25, 95HD decrease of 7.21 \& 6.04, and ASD decrease of 3.02 \& 6.51 at 10\% labeling ratio, by successively adding GazeMix, MGP, and Gaze Loss. GazeMix copies and pastes the human gaze crop on the labeled image to the unlabeled, to enhance the model’s understanding of clinically valuable content in medical image. And MGP refines the network perception in multi-scale, and align it with human perception under the guidance of Gaze Loss. Gaze Loss $L_{gaze}$teaches the model to learn human perception and focus on clinically important region. {Specifically, after adding the GazeMix module, the Dice increased by 15.87\% and 7.16\%, while the 95HD decreased by 8.58 and 3.92, respectively for different labeling ratio averagely on both datasets. When only the MGP module was added, the Dice increased by 13.47\% and 6.25\%, and the 95HD decreased by 4.36 and 2.28. This also shows that the MGP module has a positive effect on improving the model’s segmentation accuracy, and the increase is slightly smaller compared with the GazeMix module. When both GazeMix and MGP were used simultaneously, the Dice increased by 16.35\% and 7.96\%, and the 95HD decreased by 9.10 and 5.56, achieving a higher level of improvement in segmentation accuracy than using either module alone. This further highlights the enhancement of the model performance by the synergistic effect of the two modules. Based on these experimental results, it can be also verified that the GazeMix promotes a relatively higher improvement in HG-DTGL, which plays a important role in improving the model’s understanding of clinically valuable content in medical images and the segmentation accuracy.
}

\begin{table}[b]
\vspace{-0.3in}
\setlength{\tabcolsep}{2.5pt}
\centering
\small
\caption{Comparison of fixations and rapid scanning in HG-DTGL.}\label{abs_gazepoint}

\resizebox{\textwidth}{!}{
\begin{tabular}{c|c| c c c c| c c c c}
\hline
\multirow{2}*{Gaze Type} & \multirow{2}{*}{\begin{tabular}[c]{@{}c@{}}Labeling\\ Ratio\end{tabular}} & \multicolumn{4}{c|}{ACDC} & \multicolumn{4}{c}{CAMUS}\\ 
 
\cline{3-10} ~ &   & Dice$\uparrow$ & Jaccard$\uparrow$ & 95HD$\downarrow$ & ASD$\downarrow$ & Dice$\uparrow$ & Jaccard$\uparrow$ & 95HD$\downarrow$ & ASD$\downarrow$ \\

\hline
Rapid Scanning & \multirow{2}*{5\%} &86.19 &77.42 &5.32 &2.75 &83.93 &72.89 &12.24 &14.93\\
\textbf{Fixations} & &  \textbf{89.37} &\textbf{81.26} &\textbf{1.45} &\textbf{0.39} &\textbf{86.41} &\textbf{77.19} &\textbf{9.27} &\textbf{9.87}\\

\hline
Rapid Scanning & \multirow{2}*{10\%} &88.20 &79.58 &3.51 &2.43 &85.29 &75.36 &10.71 &10.46\\

\textbf{Fixations} & &  \textbf{90.87} & \textbf{83.67} & \textbf{2.07} & \textbf{0.69} & \textbf{88.05} & \textbf{79.51} & \textbf{8.22} & \textbf{7.32}\\

\hline
\end{tabular}
}
%\vspace{-0.2in}
\end{table}

{\textbf{b) Performance Analysis of GazeMix \textit{vs.} GCA \textit{vs.} Mixup \textit{vs.} CutMix:}
As shown in Table \ref{abs_mix},  the proposed GazeMix achieves the best performance, and gains average Dice increase of 2.38\% \& 1.48\%, and average 95HD decrease of 5.61 \& 4.60 with 5\% labeling ratio for ACDC and CAMUS, as well as average Dice increase of 2.59\% \& 1.85\%, and average 95HD decrease of 3.12 \& 4.32, at 10\% labeling ratio, compared with Mixup \cite{zhang2017mixup}, CutMix\cite{yun2019cutmix} and GCA\cite{GCA}. Mixup generates new image by stacking pixels on top of each other, and medical images usually have the similar spatial structure and the targets have lower contrast compared with background, which makes the network hard to distinguish the targets from the background in mixed images. And CutMix stitches randomly cropped regions from the foreground to the background image, although it alleviates the problem that the model is difficult to distinguish because of pixel mixing to some extent. However, due to its random cutting, there are problems of cutting to irrelevant areas and splicing mismatch after cutting. Compared with the other two methods, GCA and our GazeMix generate new data based on gaze. GCA crops out the radiologist non-attention region, the non-attention regions often contain a large amount of unstructured background elements. The generated data leads to weak generalization of the model due to insufficient diversity.}
Furthermore, as shown in Figure \ref{mix_type} and \ref{mix_camus}, GazeMix also enables the best visual segmentation effect with accurate structure and details, while lost RV (\textit{Figure \ref{mix_type}(b1) and (c1)}), wrong distributed RA (\textit{Figure \ref{mix_camus}(b1), (b2), (b6) and (c2)}), underestimated LV shape (\textit{Figure.\ref{mix_camus}(b1), (b4) and (c3)}), overestimated endocardium (\textit{Figure \ref{mix_camus}(b3) and (c3)}), broken LV (\textit{Figure \ref{mix_type}(b3), Figure \ref{mix_camus}(b2), (c2) and (c4)}) / RV(\textit{Figure.\ref{mix_type}(b2), (b3), (c3), (c4) and (c6)}) / myocardium (\textit{Figure \ref{mix_camus}(b2), (b3) and (c3)}) / endocardium (\textit{Figure \ref{mix_camus}(b1)-(b4), (c2) and (c4)}), and worse details remarkably exist in other methods. 
The proposed GazeMix is good at generating reliable and rich data based on the gaze guidance of human perception, so that enhances the segmentation ability of the model, even with only a small amount of labeled training data.

\begin{table}[b]
\vspace{-0.2in}
\setlength{\tabcolsep}{2.5pt}
\centering
\small
\caption{Comparison of nnUNet and gaze-based version nnUNet$_{gaze}$ on ACDC and CAMUS datasets.}\label{abs_nnUNet}

\resizebox{\textwidth}{!}{
\begin{tabular}{c| c| c c c c | c c c c}
\hline
\multirow{2}*{Architecture}  & \multirow{2}*{Variate} & \multicolumn{4}{c|}{ACDC} & \multicolumn{2}{c}{CAMUS}\\ 
 
\cline{3-10} ~&  & Dice$\uparrow$ & Jaccard$\uparrow$ & 95HD$\downarrow$ & ASD$\downarrow$ & Dice$\uparrow$ & Jaccard$\uparrow$ & 95HD$\downarrow$ & ASD$\downarrow$ \\
\hline
\multirow{2}*{nnUNet} & \textit{w/o} gaze &92.27 &86.45 &4.79 &0.79 &88.05 &82.71 &6.23 &8.58\\

 & \textbf{\textit{w} gaze} &\textbf{93.19} &\textbf{87.61} &\textbf{3.64} &\textbf{0.54} &\textbf{90.26} &\textbf{84.05} &\textbf{5.48} &\textbf{6.81}\\

\hline
\end{tabular}
}
%\vspace{-0.2in}
\end{table}

\begin{figure}[tp]
    \centering
    \includegraphics[width=0.92\linewidth]{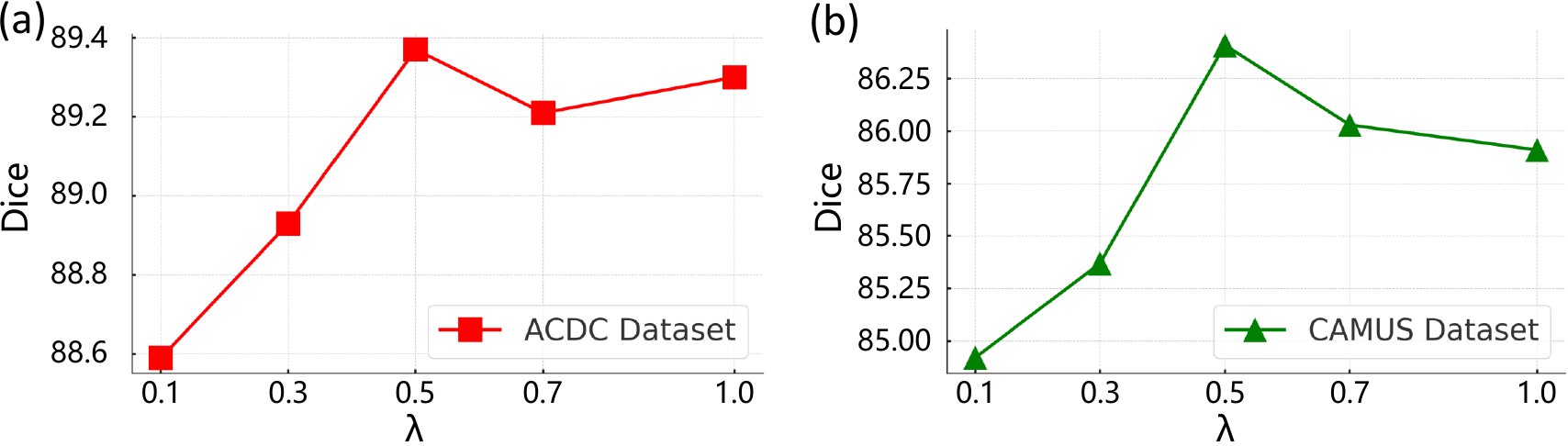}
    \vspace{-0.11in}
    \caption{Ablation study of the weights $\lambda$ in the loss function, model performs the best performance in both ACDC and CAMUS datasets with $\lambda = 0.5$ .}
\label{loss_line}
\vspace{-0.16in}
\end{figure}

\textbf{c) Performance Analysis of MGP \textit{vs.} Spatial Attention \textit{vs.} CBAM:}
As shown in Table \ref{abs_attention}, our MGP for extracting and refining network perception gaze gains the best performance on both $5\%$ and $10\%$ labeled datasets of ACDC and CAMUS, compared with situation of using spatial attention mechanism \cite{hu2018gather} and CBAM \cite{woo2018cbam} instead.   
The MGP refines multi-scale information from network perception. So that it is good at perceiving both global and local content in medical image, finely covering each tissue and facilitating network adaptively aligned to human perception, with working together in gaze teacher.

\textbf{d) Loss Weight $\lambda$ Selection:} 
We select $\lambda$ = 0.5 as loss weight value, and it effectively balances losses in training to achieve the best performance.
We set $\lambda$ in different values to reveal the influence of Gaze Loss on the model. Figure \ref{loss_line} shows the performance of the ACDC and CAMUS datasets under the $\lambda$ values 0.1, 0.3, 0.5, 0.7 and 1, and $\lambda$=0.5 suppresses other setting. Smaller $\lambda$ impedes the effect of Gaze Loss, and higher value disturbs model learning and understanding what content gazed on. 

\begin{table}[b]
\vspace{-0.2in}
\setlength{\tabcolsep}{2.5pt}
\centering
\small
\caption{The robustness studies on ACDC and CAMUS datasets.}\label{abs_robust}

\resizebox{\textwidth}{!}{
\begin{tabular}{c|c| c c c c| c c c c}
\hline
\multirow{2}*{Gaze Type} & \multirow{2}{*}{\begin{tabular}[c]{@{}c@{}}Labeling\\ Ratio\end{tabular}} & \multicolumn{4}{c|}{ACDC} & \multicolumn{4}{c}{CAMUS}\\ 
 
\cline{3-10} ~&  & Dice$\uparrow$ & Jaccard$\uparrow$ & 95HD$\downarrow$ & ASD$\downarrow$ & Dice$\uparrow$ & Jaccard$\uparrow$ & 95HD$\downarrow$ & ASD$\downarrow$ \\
\hline
Original & \multirow{2}*{5\%} & 88.09 &79.93 &2.46 &0.83 &85.56 &76.82 &10.74 &10.35\\

\textbf{Filtered} & &  \textbf{89.37} &\textbf{81.26} &\textbf{1.45} &\textbf{0.39} &\textbf{86.41} &\textbf{77.19} &\textbf{9.27} &\textbf{9.87}\\

\hline
Original & \multirow{2}*{10\%} & 89.26 &81.93 &2.32 &0.92 &86.86 &79.16 &9.26 &8.67\\

\textbf{Filtered} & &  \textbf{90.87} & \textbf{83.67} & \textbf{2.07} & \textbf{0.69} & \textbf{88.05} & \textbf{79.51} & \textbf{8.22} & \textbf{7.32}\\

\hline
\end{tabular}
}
%\vspace{-0.2in}
\end{table} 

{\textbf{e) Performance Analysis of rapid scanning vs. fixations:} As shown in Table \ref{abs_gazepoint}, the proposed method utilizes fixations, so that achieves Dice 89.37\% \& 86.41\%, 95HD 1.45 \& 9.27 with 5\% labeled training set for ACDC and CAMUS, as well as Dice 90.87\% \& 88.05\%, 95HD 2.07 \& 8.22 with 10\% labeling ratio. Compared with rapid scanning, fixations increase the Dice coefficient by approximately 2.77\% and decrease the 95HD by about 2.69 on average. The reason is that rapid scanning is mostly a simple browsing behavior, and it may introduce some transient gaze trajectories that are irrelevant to the segmentation task. This may mislead to crop the incorrect patches for mix. Moreover, the rapid scanning also cause the model to mistakenly focus on the background area excessively, thus reducing the performance. Fixation data avoids these problems and enables relevant methods to crop and focus on the correct target regions more effectively, thus improves the segmentation performance of the model.}

{\textbf{f) Analysis of Gaze Value for the Existing Architecture:} We used the fully supervised setting and implemented it on the classic architecture nnUNet. 
We trained the basic nnUNet model and its gaze-based version which introduced gaze by using the GazeMix method to perform data mixing. 
As shown in Table \ref{abs_nnUNet}, compared with the direct training nnUNet on ACDC and CAMUS, the gaze-based version nnUNet$_{gaze}$ has achieved the Dice 0.92\% improvement and the 95HD 1.15 decrease in ACDC dataset, as well as the Dice 2.21\% improvement and the 95HD 0.75 decrease in CAMUS dataset. In our exploration, due to the utilization of gaze, the performance of nnUNet$_{gaze}$ has been improved, demonstrating the positive impact and potential value of gaze on general segmentation architectures.}

{\textbf{g) The Robustness Studies:} During the process of collecting gaze data, radiologists may experience line-of-sight deviations, resulting in the generation of misleading gaze data. We further made the experiment that trained our method using the original gaze data, where the noise in the original data was not filtered. As shown in Table \ref{abs_robust}, by retaining the possible line-of-sight deviation noise generated by radiologists during the collection process, we verified the robustness of our method. Compared with the original data, the Dice of the filtered gaze data increased by 1.45\% on the ACDC dataset, and the 95HD decreased by 0.63. On the CAMUS dataset, the Dice of the filtered gaze data increased by 1.02\%, and the 95HD decreased by 1.26. This phenomenon indicates that we have effectively filtered the noise. And the segmentation performance of the original data is still within an acceptable range, reflecting the robustness of our method.}

\section{Conclusion and Discussion}
\label{sec:conclusion}
In this paper, the HG-DTGL, a novel human gaze-based semi-supervised medical image segmentation model, has been proposed. We design a novel approach GazeMix for data augmentation,
which ensures the validity of data after copy-paste and reduces the redundancy of invalid data. 
On this basis, we have specially designed a MGP module to capture the perception of the network and perform Gaze Loss with human gaze to further constrain the network focusing on human perception regions. 
Experiments on ACDC and CAMUS datasets show the superiority of the proposed method, with small gap compared with the fully-supervised segmentation model.
It thus has significant application value for the massive amount of unlabeled data in medical field.

{Although the HG-DTGL model has achieved remarkable results through innovative data augmentation and perception constraint mechanisms, the complex visual environment in medical scenarios poses special impacts to deep learning models based on gaze data. Since radiologists need to quickly switch their areas of focus during clinical diagnosis, the original gaze data inherently contains a large number of transient gaze trajectories that are irrelevant to semantic segmentation. This noisy characteristic may weaken the effectiveness of the gaze data. Regarding this issue, Wang et al.\cite{wang2022follow} proposed using homoscedastic uncertainty to model visual attention and suppress inter-observer variability, so that promoted gaze information learning and facilitated computer-aided diagnosis (CAD) of osteoarthritis.  
Zhao et al.\cite{zhao2024mining} adopted a confidence score to deal with the noise in gaze data, so that used to guide contrastive pre-training for CAD. 
Following the previous work \cite{xie2024integrating}, gaze points are divided into fixation points and saccade points in our study, and the fixations with spatial information are used for medical segmentation model training. Our robustness study shows that even when using unfiltered gaze data, the performance of our method in the semi-supervised medical image segmentation task is still acceptable. In the future, by addressing the noise issue in gaze data, we hope to further enhance the performance and applicability of this method.}

\section*{CRediT authorship contribution statement}
\textbf{Rongjun Ge:} Writing–original draft, Software, Methodology, Conceptualization, Funding acquisition. \textbf{Chong Wang:} Validation, Software, Formal analysis. \textbf{Yuting He:} Conceptualization, Formal analysis. \textbf{Yuxin Liu:} Validation. \textbf{Chunqiang Lu:} Formal analysis. \textbf{Cong Xia:} Resources. \textbf{Yehui Jiang:} Resources. \textbf{Fangyi Xu:} Resources. \textbf{Yinsu Zhu:} Resources. \textbf{Daoqiang Zhang:} Supervision, Funding acquisition. \textbf{Chenyu Liu:} Conceptualization, Funding acquisition. \textbf{Yang Chen:} Supervision, Resources, Funding acquisition. 
{\textbf{Shuo Li:} Conceptualization, Supervision.}

\section*{Declaration of competing interest}
The authors declare that they have no known competing financial interests or personal relationships that could have appeared to influence the work reported in this paper.

\section*{Acknowledgments}
This work was supported in part by the National Natural Science Foundation of China under Grant 62101249, Grant T2225025 and Grant 62136004; in part by the Natural Science Foundation of Jiangsu Province under Grant BK20210291; in part by the Fundamental Research Funds for the Central Universities under grant 226-2024-00185.

\bibliographystyle{elsarticle-num}
\bibliography{ref2}

\end{document}